UDC: 538.9 Condensed matter Physics, Solid state Physics, Theoretical Condensed matter Physics

# MAGNETIC ENERGY OF SC FERROMAGNETIC FILMS WITH THREE LAYERS AS DESCRIBED BY THIRD ORDER PERTURBED HEISENBERG HAMILTONIAN

P. Samarasekara[1*] and N.U.S. Yapa[2]

[1]Department of Physics, University of Peradeniya, Peradeniya, Sri Lanka

[2]Department of Physics, Open University of Sri Lanka, Kandy, Sri Lanka

***Abstract***

*The solution of third order perturbed Heisenberg Hamiltonian of simple cubic ferromagnetic ultra-thin films with three layers were found. All the magnetic energy parameters such as spin exchange interaction, magnetic dipole interaction, second order magnetic anisotropy, fourth order magnetic anisotropy, applied magnetic field, demagnetization factor and stress induced anisotropy were included in the third order perturbed Heisenberg Hamiltonian. 3-D plots of stress induced anisotropy, out of plane magnetic field, demagnetization factor and spin exchange interaction are presented in this manuscript. Magnetic easy and hard directions were determined using these 3-D plots. MATLAB program was employed to solve the equation with seven parameters.*

## 1. Introduction:

Ferromagnetic films have been described using many different theoretical models. EuTe films with surface elastic stresses have been theoretically investigated using Heisenberg Hamiltonian [1]. Magnetostriction of dc magnetron sputtered FeTaN thin films has been theoretically explained using the theory of De Vries [2]. Magnetic layers of Ni on Cu have been theoretically studied using the Korringa-Kohn-Rostoker Green's function method [3]. Electric and magnetic properties of multiferroic thin films have been theoretically investigated by modified Heisenberg and transverse Ising model using Green's function technique [4]. The quasistatic magnetic hysteresis of ferromagnetic thin films grown on a vicinal substrate has been theoretically described by Monte Carlo simulations within a 2D model [5]. Structural and magnetic properties of two dimensional FeCo orders alloys deposited on W(110) substrates have been investigated using first principles band structure theory [6].

Heisenberg Hamiltonian was employed to solve the problems of magnetic thin films as summarized below. Ferromagnetic thin films have been previously studied using the Heisenberg Hamiltonian with spin exchange interaction, magnetic dipole interaction, applied magnetic field, second and fourth order magnetic anisotropy [7, 8, 9]. Magnetization reversal and domain structure in thin magnetic films have been theoretically investigated [10]. In-plane dipole coupling anisotropy of a square ferromagnetic Heisenberg monolayer has been described using Heisenberg Hamiltonian [11]. Effect of the interracial coupling on the magnetic ordering in ferro-antiferromagntic bilayers has been studied using Heisenberg Hamiltonian [12].

Previously strontium ferrite [13] and nickel ferrite [14] films were synthesized using sputtering by us. In addition, lithium mixed ferrite films were fabricated using pulsed laser deposition [15]. For all these films, the coercivity of film increased due to the stress induced anisotropy. The change of coercivity due to the stress induced anisotropy was qualitatively calculated for all these films. The calculated values of the change of coercivity agreed with the experimentally found values. So the stress induced anisotropy plays a major role in magnetic thin fabrications. Previously the Heisenberg Hamiltonian was employed to investigate the second order perturbed energy of ultrathin ferromagnetic films [16], thick ferromagnetic films [17], unperturbed energy of spinel ferrite films [18], second order perturbed energy of thick ferromagnetic films [19], third order perturbed energy of thick spinel ferrite [20], third order perturbed energy of thin spinel ferrite [21], second order perturbed energy of spinel ferrite [22] and spin reorientation of barium ferrite [23].

**2. Model:**

Modified classical Heisenberg Hamiltonian with all the magnetic parameters is given as below.

$$H = -\frac{J}{2}\sum_{m,n}\vec{S}_m.\vec{S}_n + \frac{\omega}{2}\sum_{m\neq n}\left(\frac{\vec{S}_m.\vec{S}_n}{r_{mn}^{\ 3}} - \frac{3(\vec{S}_m.\vec{r}_{mn})(\vec{r}_{mn}.\vec{S}_n)}{r_{mn}^{\ 5}}\right) - \sum_m D_{\lambda_m}^{\ (2)}(S_m^{\ z})^2 - \sum_m D_{\lambda_m}^{\ (4)}(S_m^{\ z})^4$$

$$-\sum_{m,n}[\vec{H} - (N_d\vec{S}_n/\mu_0)].\vec{S}_m - \sum_m K_s Sin2\theta_m \qquad (1)$$

Here J is spin exchange interaction, $\omega$ is the strength of long range dipole interaction, $\theta$ is azimuthal angle of spin, $D_m^{(2)}$ and $D_m^{(4)}$ are second and fourth order anisotropy constants, $H_{in}$ and $H_{out}$ are in plane and out of plane applied magnetic fields, $K_s$ is stress induced anisotropy constant, n and m are spin plane indices, and N is total number of layers in film. When the stress applies normal to the film plane, the angle between m$^{th}$ spin and the stress is $\theta_m$. Because the size of the unit cell of ferromagnetic structure is taken as one, $r_{mn}$ is taken as a fraction of the size of the unit cell of ferromagnetic structure.

The total energy per unit spin can be deduced to the following equation.

$$E(\theta)= -\frac{1}{2}\sum_{m,n=1}^{N}[(JZ_{|m-n|}-\frac{\omega}{4}\Phi_{|m-n|})\cos(\theta_m-\theta_n)-\frac{3\omega}{4}\Phi_{|m-n|}\cos(\theta_m+\theta_n)]$$

$$-\sum_{m=1}^{N}(D_m^{(2)}\cos^2\theta_m+D_m^{(4)}\cos^4\theta_m+H_{in}\sin\theta_m+H_{out}\cos\theta_m)$$

$$+\sum_{m,n=1}^{N}\frac{N_d}{\mu_0}\cos(\theta_m-\theta_n)-K_s\sum_{m=1}^{N}\sin 2\theta_m \qquad (2)$$

where m (or n) $Z_{|m-n|},\phi_{|m-n|},\theta_m (or\ \theta_n)$, *N, $H_{in}$* and *$H_{out}$* being indices of layers, number of nearest spin neighbors, constant arising from summation of dipole interactions, azimuthal angles of spins, total number of layers, in plane applied field and out of plane applied field, respectively. With some perturbation, above angles $\theta_m$ and $\theta_n$ measured with film normal can be expressed in forms of $\theta_m=\theta+\varepsilon_m$ and $\theta_n=\theta+\varepsilon_n$, and above energy can be expanded up to the third order of $\varepsilon$ as following. Here $\varepsilon_m$ (or $\varepsilon_n$) is a small perturbation of the angle.

$$E(\theta)=E_0+E(\varepsilon)+E(\varepsilon^2)+E(\varepsilon^3) \qquad (3)$$

Here $E_0= -\frac{1}{2}\sum_{m,n=1}^{N}(JZ_{|m-n|}-\frac{\omega}{4}\Phi_{|m-n|})+\frac{3\omega}{8}\cos 2\theta\sum_{m,n=1}^{N}\Phi_{|m-n|}$

$$-\cos^2\theta\sum_{m=1}^{N}D_m^{(2)}-\cos^4\theta\sum_{m=1}^{N}D_m^{(4)}-N(H_{in}\sin\theta+H_{out}\cos\theta-\frac{N_d}{\mu_0}+K_s\sin 2\theta)$$

$$E(\varepsilon)=-\frac{3\omega}{8}\sin 2\theta\sum_{m,n=1}^{N}\Phi_{|m-n|}(\varepsilon_m+\varepsilon_n)+\sin 2\theta\sum_{m=1}^{N}D_m^{(2)}\varepsilon_m+2\cos^2\theta\sin 2\theta\sum_{m=1}^{N}D_m^{(4)}\varepsilon_m$$

$$-H_{in}\cos\theta\sum_{m=1}^{N}\varepsilon_m+H_{out}\sin\theta\sum_{m=1}^{N}\varepsilon_m-2K_s\cos 2\theta\sum_{m=1}^{N}\varepsilon_m$$

$$E(\varepsilon^2) = \frac{1}{4}\sum_{m,n=1}^{N}(JZ_{|m-n|} - \frac{\omega}{4}\Phi_{|m-n|})(\varepsilon_m - \varepsilon_n)^2 - \frac{3\omega}{16}\cos 2\theta \sum_{m,n=1}^{N}\Phi_{|m-n|}(\varepsilon_m + \varepsilon_n)^2$$

$$-(\sin^2\theta - \cos^2\theta)\sum_{m=1}^{N} D_m^{(2)}\varepsilon_m^{\ 2} + 2\cos^2\theta(\cos^2\theta - 3\sin^2\theta)\sum_{m=1}^{N} D_m^{(4)}\varepsilon_m^{\ 2}$$

$$+\frac{H_{in}}{2}\sin\theta\sum_{m=1}^{N}\varepsilon_m^{\ 2} + \frac{H_{out}}{2}\cos\theta\sum_{m=1}^{N}\varepsilon_m^{\ 2} - \frac{N_d}{2\mu_0}\sum_{m,n=1}^{N}(\varepsilon_m - \varepsilon_n)^2$$

$$+2K_s \sin 2\theta \sum_{m=1}^{N}\varepsilon_m^{\ 2}$$

$$E(\varepsilon^3) = \frac{\omega}{16}\sin 2\theta \sum_{m,n=1}^{N}(\varepsilon_m + \varepsilon_n)^3 \phi_{|m-n|} - \frac{4}{3}\cos\theta\sin\theta\sum_{m=1}^{N} D_m^{(2)}\varepsilon_m^{\ 3}$$

$$-4\cos\theta\sin\theta(\frac{5}{3}\cos^2\theta - \sin^2\theta)\sum_{m=1}^{N} D_m^{(4)}\varepsilon_m^{\ 3} + \frac{H_{in}}{6}\cos\theta\sum_{m=1}^{N}\varepsilon_m^{\ 3}$$

$$-\frac{H_{out}}{6}\sin\theta\sum_{m=1}^{N}\varepsilon_m^{\ 3} + \frac{4K_s}{3}\cos 2\theta\sum_{m=1}^{N}\varepsilon_m^{\ 3} \qquad (4)$$

After using the constraint $\sum_{m=1}^{N}\varepsilon_m = 0$, $E(\varepsilon) = \vec{\alpha}.\vec{\varepsilon}$

Here $\vec{\alpha}(\varepsilon) = \vec{B}(\theta)\sin 2\theta$ are the terms of matrices with

$$B_\lambda(\theta) = -\frac{3\omega}{4}\sum_{m=1}^{N}\Phi_{|\lambda-m|} + D_\lambda^{(2)} + 2D_\lambda^{(4)}\cos^2\theta \qquad (5)$$

Also $E(\varepsilon^2) = \frac{1}{2}\vec{\varepsilon}.C.\vec{\varepsilon}$ (6)

Here the elements of matrix C can be given as following,

$$C_{mn} = -(JZ_{|m-n|} - \frac{\omega}{4}\Phi_{|m-n|}) - \frac{3\omega}{4}\cos 2\theta\Phi_{|m-n|} + \frac{2N_d}{\mu_0}$$

$$+\delta_{mn}\{\sum_{\lambda=1}^{N}[JZ_{|m-\lambda|} - \Phi_{|m-\lambda|}(\frac{\omega}{4} + \frac{3\omega}{4}\cos 2\theta)] - 2(\sin^2\theta - \cos^2\theta)D_m^{(2)}$$

$$+4\cos^2\theta(\cos^2\theta - 3\sin^2\theta)D_m^{(4)} + H_{in}\sin\theta + H_{out}\cos\theta - \frac{4N_d}{\mu_0} + 4K_s\sin 2\theta\} \qquad (7)$$

In addition, third order can be expressed as the

$$E(\varepsilon^3) = \varepsilon^2\beta.\vec{\varepsilon} \qquad (8)$$

Here elements of matrix β can be given as following,

$$\beta_{mn} = \frac{3\omega}{8}\sin 2\theta\Phi_{|m-n|} + \delta_{mn}\{\frac{\omega}{8}\sin 2\theta[A_m - \Phi_0] - \frac{4}{3}\cos\theta\sin\theta D_m^{(2)}$$

$$-4\cos\theta\sin\theta(\frac{5}{3}\cos^2\theta - \sin^2\theta)D_m^{(4)} + \frac{H_{in}}{6}\cos\theta - \frac{H_{out}}{6}\sin\theta$$

$$+\frac{4K_s}{3}\cos 2\theta\}$$

Also $\beta_{nm}=\beta_{mn}$, implying that matrix β is symmetric.

After substituting equations (8) and (6) in equation (3), total energy can be expressed as

$$E(\theta)=E_0+\vec{\alpha}.\vec{\varepsilon} + \frac{1}{2}\vec{\varepsilon}.C.\vec{\varepsilon} + \varepsilon^2\beta.\vec{\varepsilon}$$

At the energetically favorable state, the derivative of above E(θ) with respect to ε will be zero. Using that condition, ε can be found. After substituting that ε in above equation of E(θ), following equation can be derived.

$$E(\theta)=E_0 - \frac{1}{2}\vec{\alpha}.C^+.\vec{\alpha} - (C^+\alpha)^2\vec{\beta}(C^+\alpha)$$

The total magnetic energy have been calculated only for three layers ($N$=3), and the equation have been proved under the assumption of $D_1^{(2)}=D_2^{(2)}=D_3^{(2)}$ and $D_1^{(4)}=D_2^{(4)}=D_3^{(4)}$.

Following equation has been used to calculate the elements of matrix $C^+$.

$$C.C^+ = 1 - \frac{E}{N}$$

Each element of matrix $E$ is one, and $C^+$ is a pseudo inverse.

**3. Results and discussion:**

For ferromagnetic films with simple cubic structure, $Z_0$=4,$Z_1$=1,$Z_2$=0 and $\Phi_0 = 9.0336$, $\Phi_1 = -0.3275$ . All the data are given here for film films with three layers (N=3). Previously, the third order perturbed Heisenberg Hamiltonian with only the magnetic exchange energy, second order anisotropy, and the stress induced anisotropy terms has been solved by us [24]. Here all the seven magnetic energy terms will be considered.

Figure 1 shows the 3-D plot of $\frac{E(\theta)}{\omega}$ versus angle and $\frac{K_s}{\omega}$. Magnetic easy directions are observed at $\frac{K_s}{\omega}$=11, 21, ----etc. Hard directions appear at $\frac{K_s}{\omega}$=7, 17, 27, -----etc. Two magnetic hard directions with different energies can be observed. Similarly two magnetic easy directions with different energies can be seen. The $\frac{K_s}{\omega}$ values at hard directions with higher energy and easy directions with lower energy are given here. Hard and easy directions can be observed at 8 and 4 radians, respectively. Other parameters were kept at $\frac{J}{\omega}=\frac{H_{in}}{\omega}=\frac{H_{out}}{\omega}=\frac{N_d}{\mu_0\omega}=10$, $\frac{D_m^{(2)}}{\omega}=30$ and $\frac{D_m^{(4)}}{\omega}=20$ for this simulation. Previously, 3-D plot of $\frac{E(\theta)}{J}$ versus angle and $\frac{K_s}{J}$ was plotted for Heisenberg Hamiltonian with only the magnetic exchange energy, second order anisotropy, fourth order anisotropy and the stress induced anisotropy terms for N=3. In that case, the 3-D plot had a less number of peaks. However, the values of $\frac{E(\theta)}{J}$ were higher than the values of $\frac{E(\theta)}{\omega}$ by an order of $10^{18}$ because the energy was divided by J and ω in previous and this case, respectively [24].

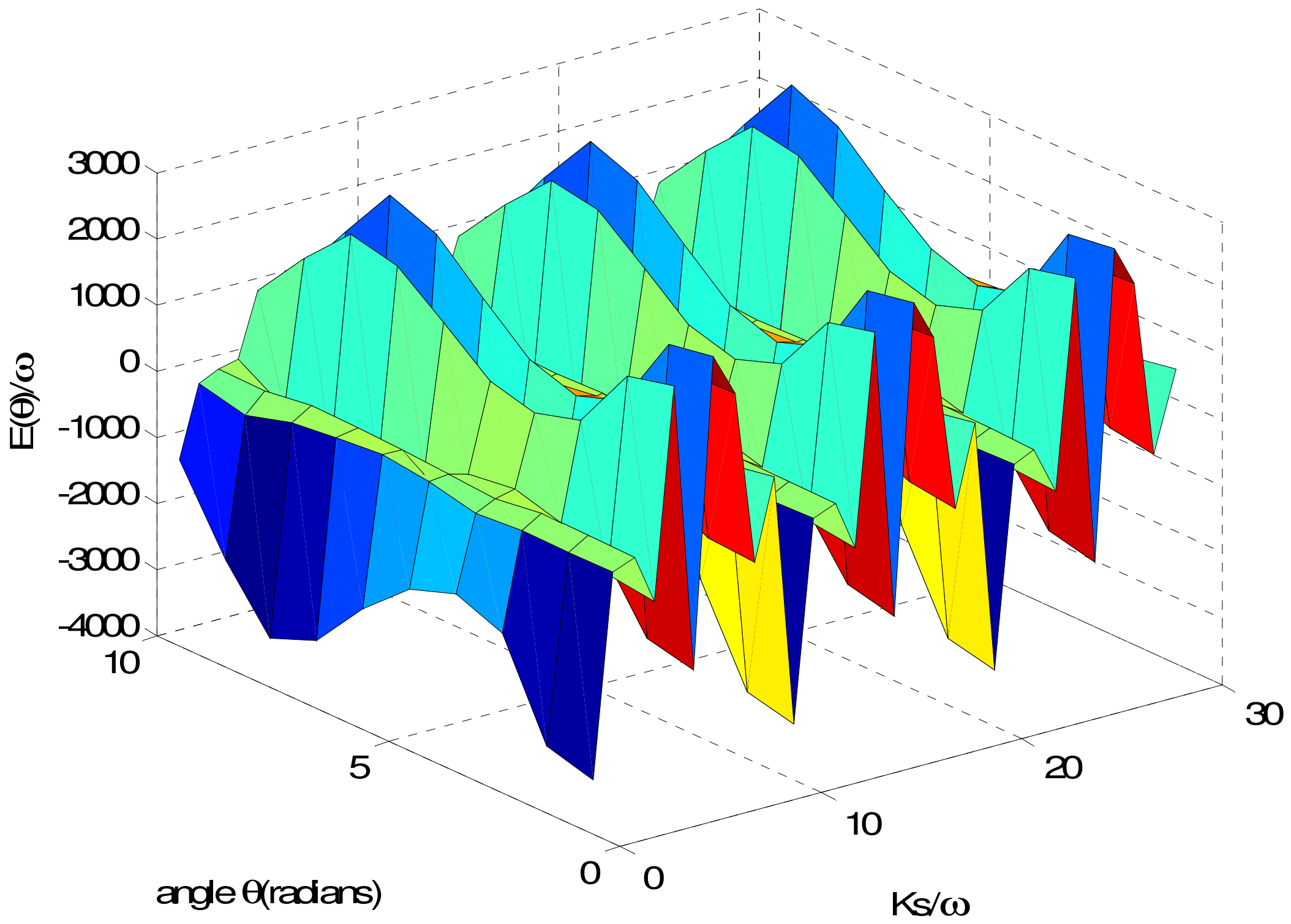

Figure 1: 3-D plot of $\frac{E(\theta)}{\omega}$ versus angle and $\frac{K_s}{\omega}$.

Figure 2 shows the 3-D plot of $\frac{E(\theta)}{\omega}$ versus angle and $\frac{H_{out}}{\omega}$. Here other parameters were kept at $\frac{J}{\omega}=\frac{H_{in}}{\omega}=\frac{K_s}{\omega}=\frac{N_d}{\mu_0\omega}=10$, $\frac{D_m^{(2)}}{\omega}=30$ and $\frac{D_m^{(4)}}{\omega}=20$ for this simulation. Magnetic easy directions can be observed at $\frac{H_{out}}{\omega}$=2, 7, 12, ---etc. Magnetic hard directions appear at $\frac{H_{out}}{\omega}$=1, 11, 21, ---etc. In addition to these major magnetic easy and hard directions, some minor easy and hard directions can be observed. Hard and easy directions of magnetizations appear at 4 and 5.57 radians, respectively. Angle between easy and hard directions is 90 degrees in this case.

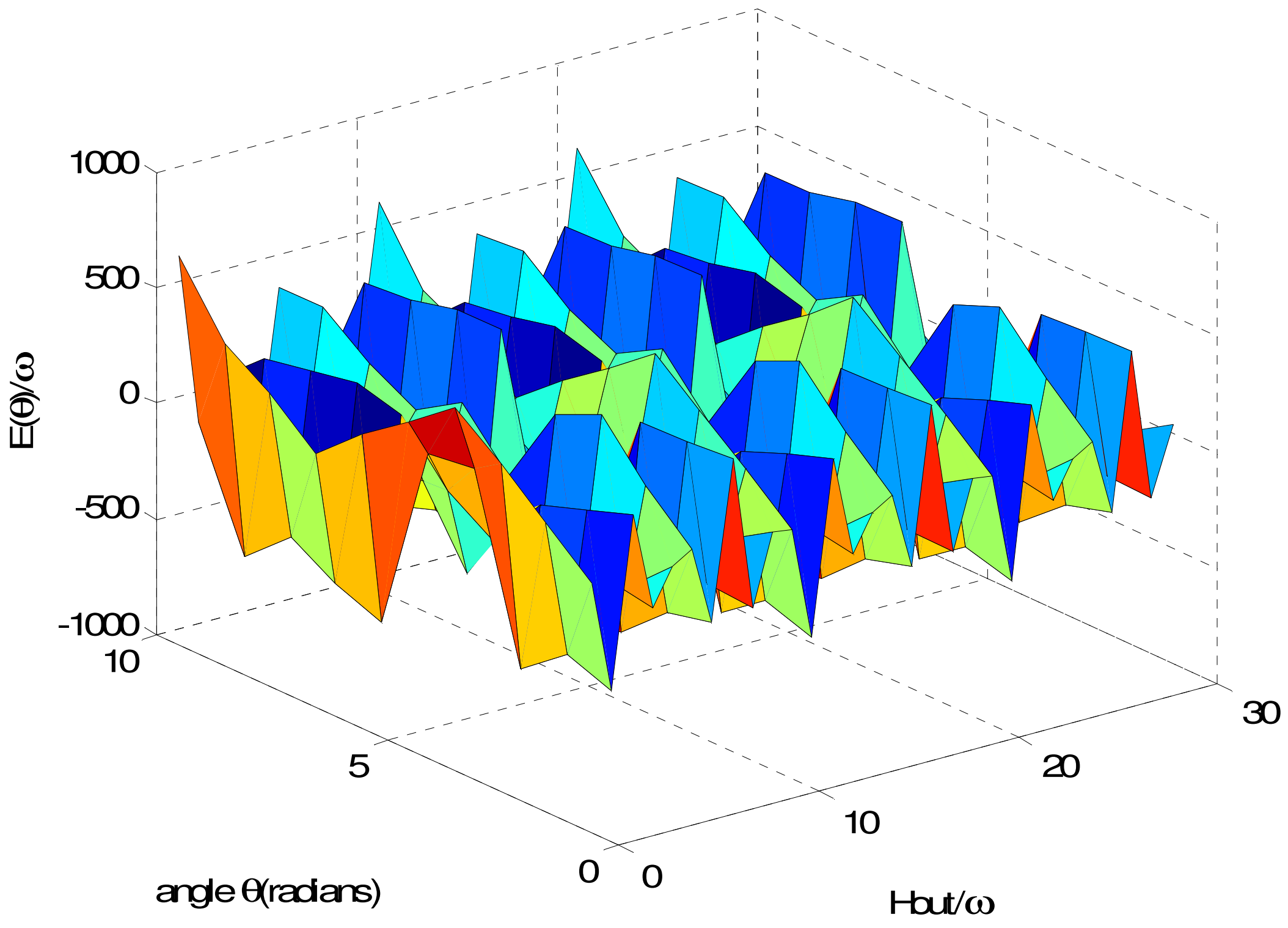

Figure 2: 3-D plot of $\frac{E(\theta)}{\omega}$ versus angle and $\frac{H_{out}}{\omega}$.

Figure 3 shows the 3-D plot of $\frac{E(\theta)}{\omega}$ versus angle and $\frac{N_d}{\mu_0\omega}$. Other parameters were kept at $\frac{J}{\omega}=\frac{H_{in}}{\omega}=\frac{K_s}{\omega}=\frac{H_{out}}{\omega}=10$, $\frac{D_m^{(2)}}{\omega}=30$ and $\frac{D_m^{(4)}}{\omega}=20$ for this simulation. Easy directions of magnetization can be observed at $\frac{N_d}{\mu_0\omega}$=1, 11, 21, ----etc. Hard directions of magnetization appear at $\frac{N_d}{\mu_0\omega}$=10, 20, 30, ----etc.

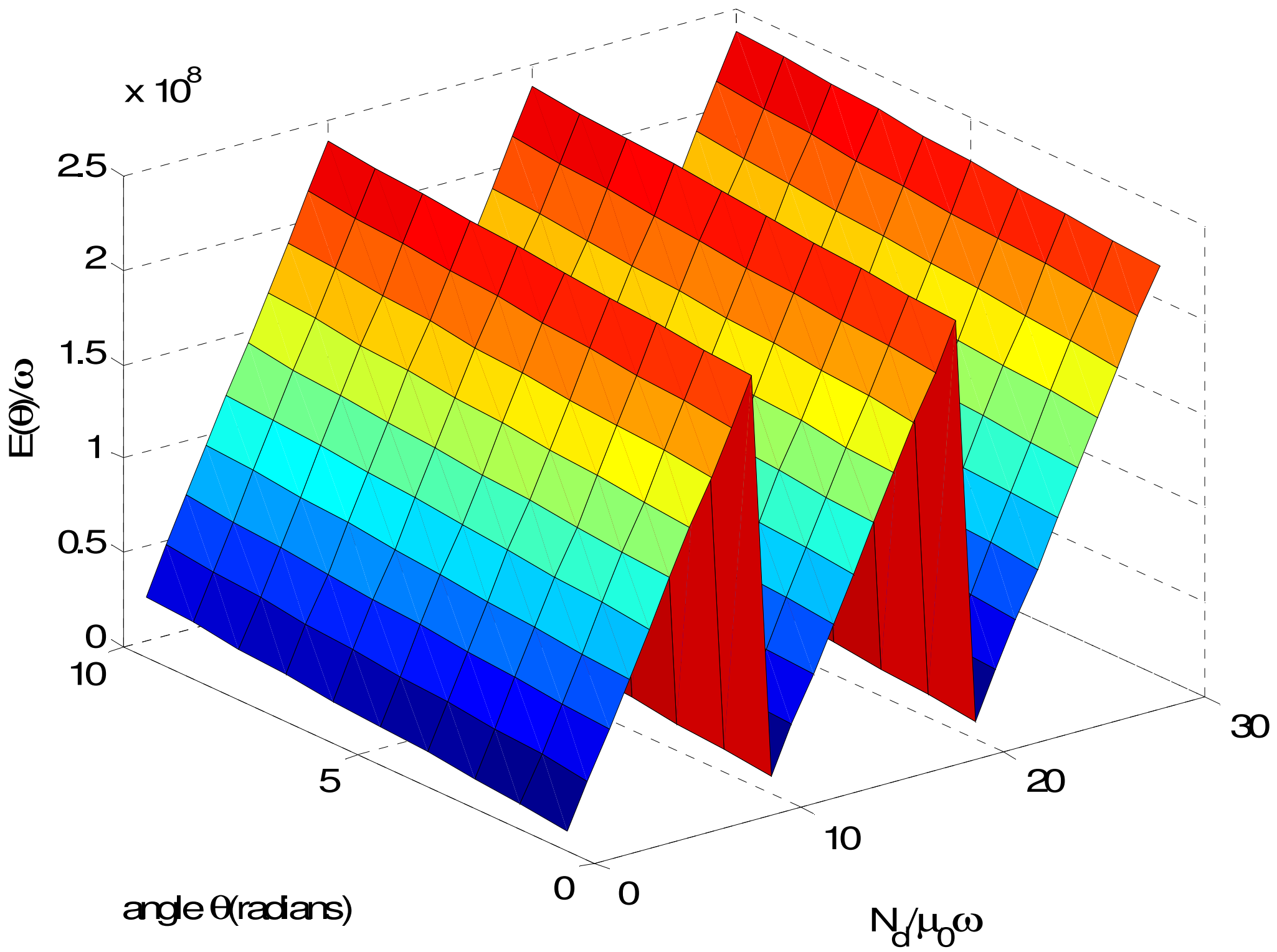

Figure 3: 3-D plot of $\frac{E(\theta)}{\omega}$ versus angle and $\frac{N_d}{\mu_0 \omega}$.

Figure 4 indicates the 3-D plot of $\frac{E(\theta)}{\omega}$ versus angle and $\frac{J}{\omega}$. Other parameters were kept at $\frac{H_{in}}{\omega} = \frac{K_s}{\omega} = \frac{H_{out}}{\omega} = \frac{N_d}{\mu_0 \omega} = 10$, $\frac{D_m^{(2)}}{\omega} = 30$ and $\frac{D_m^{(4)}}{\omega} = 20$ in this case. Easy directions can be observed at $\frac{J}{\omega}$=10, 20, 30, ----etc. Magnetic hard directions appear at $\frac{J}{\omega}$=1, 11, 21, ----etc. Hard and easy directions of magnetizations appear at 3 and 6 radians, respectively.

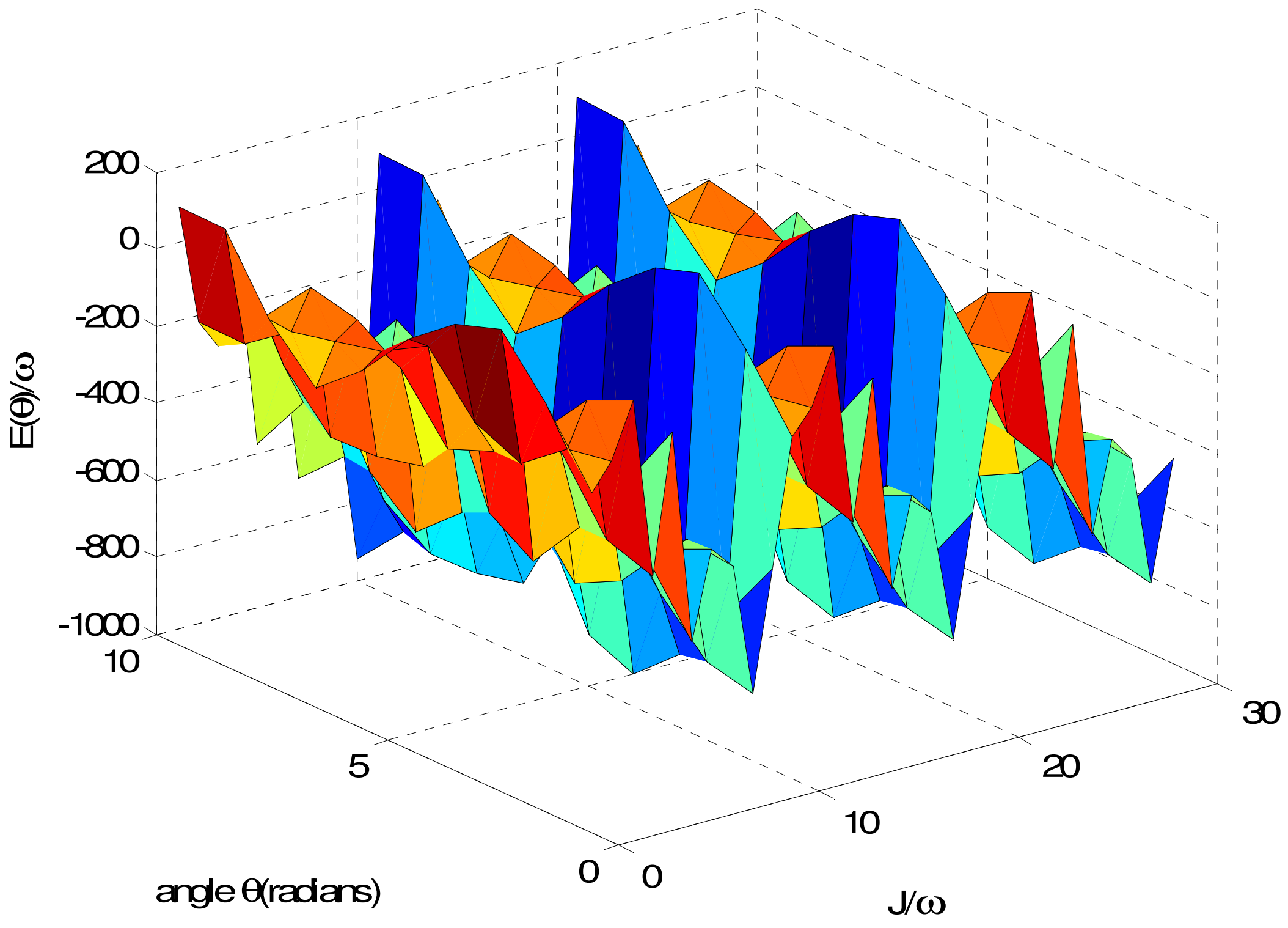

Figure 4: 3-D plot of $\frac{E(\theta)}{\omega}$ versus angle and $\frac{J}{\omega}$.

**4. Conclusion:**

Several magnetic easy and hard directions were found in the plots of $\frac{K_s}{\omega}$, $\frac{H_{out}}{\omega}$ and $\frac{J}{\omega}$. However, only few peaks were found in the plot of $\frac{N_d}{\mu_0\omega}$. This may be related to the fact that $\frac{N_d}{\mu_0\omega}$ contributes only a constant to the equation of total energy. According to figure 1, hard and easy directions can be observed at 8 and 4 radians, respectively. According to figure 2, hard and easy directions of magnetizations appear at 4 and 5.57 radians, respectively. Easy directions of magnetization can be observed at $\frac{N_d}{\mu_0\omega}$=1, 11, 21, ----etc and, hard directions of magnetization appear at $\frac{N_d}{\mu_0\omega}$=10, 20, 30, ----etc. Hard and easy directions of magnetizations appear at 3 and 6 radians, respectively according to figure 4. The angle between easy and hard directions is 90 degrees only according to figure 2.